\documentclass[aps,prx,reprint,superscriptaddress]{revtex4-2}

\usepackage[T1]{fontenc}
\usepackage[utf8]{inputenc}

\usepackage{amsmath}
\usepackage{amsfonts}
\usepackage{siunitx}
\usepackage{graphicx}
\usepackage{nicefrac}
\usepackage{multirow}
\usepackage{placeins}
\usepackage{hyperref}

\newcommand{\TheTitle}{Dynamical Heart Beat Correlations during Running}

\newcommand{\hd}{h_{+}} 
\newcommand{\hs}{h_{-}} 
\newcommand{\Fs}{\tilde{F}_t(s)} 
\newcommand{\Fsm}{\tilde{F}_{t}(s-1)} 
\newcommand{\Fsp}{\tilde{F}_{t}(s+1)} 
\newcommand{\abs}[1]{{\left|#1\right|}} 
\newcommand{\mat}[1]{{\mathbf{#1}}} 
\newcommand{\transpose}{^{\mathbf{\top}}}
\newcommand{\mean}[1]{{\left\langle #1 \right\rangle}} 
\newcommand{\dd}[1]{\mathop{\mathrm{d}#1}} 
\newcommand{\dlf}{a} 
\newcommand{\cC}{{\cal C}}
\newcommand{\RRmax}{\mathrm{RR_{\text{max}}}}
\newcommand{\RRmin}{\mathrm{RR_{\text{min}}}}
\newcommand{\RRmed}{\mathrm{RR_{\text{med}}}}
\newcommand{\lengthmed}{l_{\text{med}}}
\newcommand{\thresholdmed}{c_{\text{med}}}
 
\newcommand{\FiniteDifferenceApproximation}{\alpha(t,s) \approx \big[ &\hs^2\Fsp + \left(\hd^2-\hs^2\right)\Fs \nonumber\\%
& - \hd^2\Fsm \big] / \big[ \hs\hd\left(\hd + \hs\right) \big]}

\newcommand{\midrule}{\hline}
\newcommand{\bottomrule}{\toprule}

\begin{document}

\title{\TheTitle}

\newcommand{\affilTUNI}{Computational Physics Laboratory, Tampere University, FI-33720 Tampere, Finland}
\newcommand{\affilCNRS}{Laboratoire de Physique Th\'eorique et Mod\`eles Statistiques, CNRS UMR 8626, Universit\'e Paris-Sud, Universit\'e Paris-Saclay, 91405 Orsay cedex, France}
\newcommand{\affilMIT}{Massachusetts Institute of Technology, MultiScale Materials Science for Energy and Environment, Joint MIT-CNRS Laboratory (UMI 3466), Cambridge, Massachusetts 02139, USA}

\author{Matti Molkkari}
\email{matti.molkkari@tuni.fi}
\affiliation{\affilTUNI}

\author{Giorgio Angelotti}
\email{giorgio.angelotti@isae-supaero.fr}
\affiliation{\affilCNRS}

\author{Thorsten Emig}
\email{thorsten.emig@lptms.u-psud.fr}
\affiliation{\affilCNRS}
\affiliation{\affilMIT}

\author{Esa R\"as\"anen}
\email{esa.rasanen@tuni.fi}
\affiliation{\affilTUNI}
\affiliation{\affilCNRS}

\date{2020-07-03}

\begin{abstract}
Fluctuations of the human heart beat constitute a complex system that has been studied mostly under resting conditions using conventional time series analysis methods.
During physical exercise, the variability of the fluctuations is reduced, and the time series of beat-to-beat RR intervals (RRIs) become highly non-stationary. Here we develop a dynamical approach to analyze the time evolution of RRI correlations in running across various training and racing events under real-world conditions. 
In particular, we introduce dynamical detrended fluctuation analysis and dynamical partial autocorrelation functions, which are able to detect real-time changes in the scaling and correlations of the RRIs as functions of the scale and the lag. We relate these changes to the exercise intensity quantified by the heart rate (HR).
Beyond subject-specific HR thresholds the RRIs show multiscale anticorrelations with both universal and individual scale-dependent structure that is potentially affected by the stride frequency.
These preliminary results are encouraging for future applications of the dynamical statistical analysis in exercise physiology and cardiology, and the presented methodology is also applicable across various disciplines.
 \end{abstract}

\maketitle

\section*{Introduction\label{sec:introduction}}

The increasing popularity and accuracy of wearable devices and sensors present new opportunities to study human physiology in a continuous, non-invasive manner for a huge number of subjects under real-world conditions. These devices enable the measurement of a plethora of physiological and mechanical signals such as the heart rate, beat-to-beat (RR) intervals, overall motion via GPS, motion of specific body locations via accelerations, and skin temperature. These data can be recorded in real time, often at one second intervals, and uploaded to web services. To date, most recorded data are not analyzed in scientific rigour due to a lack of suitable models for the dynamics of physiological signals under various intensities of exercise load, and also due to restricted availability of the data (property of industry and users). This limits opportunities for a better understanding of complex physiological processes, diagnostics and monitoring for patients in rehabilitation, and the optimal training of athletes.

However, it has been long known that a variety of physiological conditions and cardiac diseases affect heart rate variability (HRV) and the correlations in RR intervals \cite{Goldberger2002}.
In exercise physiology, HRV is often used at rest to evaluate recovery, fatigue and overtraining. It is known that during exercise the overall variability of the RR intervals (RRI) is strongly suppressed. Regardless, the RRI correlations contain valuable information even during exercise \cite{Karasik2002,Hautala2003,Gronwald2019}.
For example, the possibility to determine certain physiological thresholds, such as the anaerobic threshold, from the frequency spectrum of HRV has been examined~\cite{Buchheit2007,DiMichelle2012}. Often the relative importance of low-frequency (LF: 0.04--0.15 Hz) and high-frequency (HF: 0.15--0.4 Hz) spectral power is studied during exercise.
Using this concept as a measure of the relative sympathetic (SNS) and parasympathetic nervous system (PNS) activity, it has been shown that the PNS activity decreases dramatically during exercise~\cite{Yamamoto:1991sh}.
In contrast, the SNS activity remains unchanged past the first ventilatory threshold before increasing abruptly~\cite{Yamamoto:1991sh}.
However, the use of the HF/LF ratio to measure cardiac sympatho-vagal balance has
been criticized~\cite{Billman2013}.
Moreover, it is known that Fourier decomposition of dynamic signals is often hampered by non-stationarity.

To overcome the complications of Fourier methods and non-stationarities, we base our analysis on detrended fluctuation analysis\cite{Peng1994} (DFA), which was developed to measure correlations in non-stationary time series by utilizing systematic detrending \cite{Peng1994,Peng1995,Kantelhardt2001}.
Furthermore, we are interested in analyzing real world exercises recorded with readily available commercial sports watches.
Hence, we study real-time correlations of RRIs during marathon races (group~M) and freeform training runs (group~T).
Such uncontrolled data may be plagued by severe non-stationary conditions, and the conventional division into short- and long-scale DFA exponents\cite{Peng1995,Shaffer2017} is likely to be insufficient. To this end, we introduce dynamic DFA (DDFA) for the accurate determination of time- and scale-dependent scaling exponents $\alpha(t,s)$ with high temporal resolution.
To check the consistency of our methodology, we also apply similar dynamic approach to partial autocorrelation functions (PACFs) to obtain their dynamic counterpart (DPACF).

\section*{Results\label{sec:results}}

Our main result is the discovery of scale-dependent anticorrelations ($\alpha<0.5$) in the RRIs during running that vary with the heart rate. The anticorrelations appear after the HR exceeds a subject-specific threshold. Their magnitude and the scale with the most dominant anticorrelations changes with exercise intensity.
We find that the DDFA method can reliably determine the {\it dynamic, scale-dependent} scaling exponent $\alpha(t,s)$ (please see the Supplementary Information for its numerical validation). Hence, it provides a powerful method for measuring multiscale correlations of non-stationary physiological signals. The results from the DDFA and DPACF methods are found to be mutually consistent.

\subsection*{Marathon Races}

\begin{figure*}
    \includegraphics{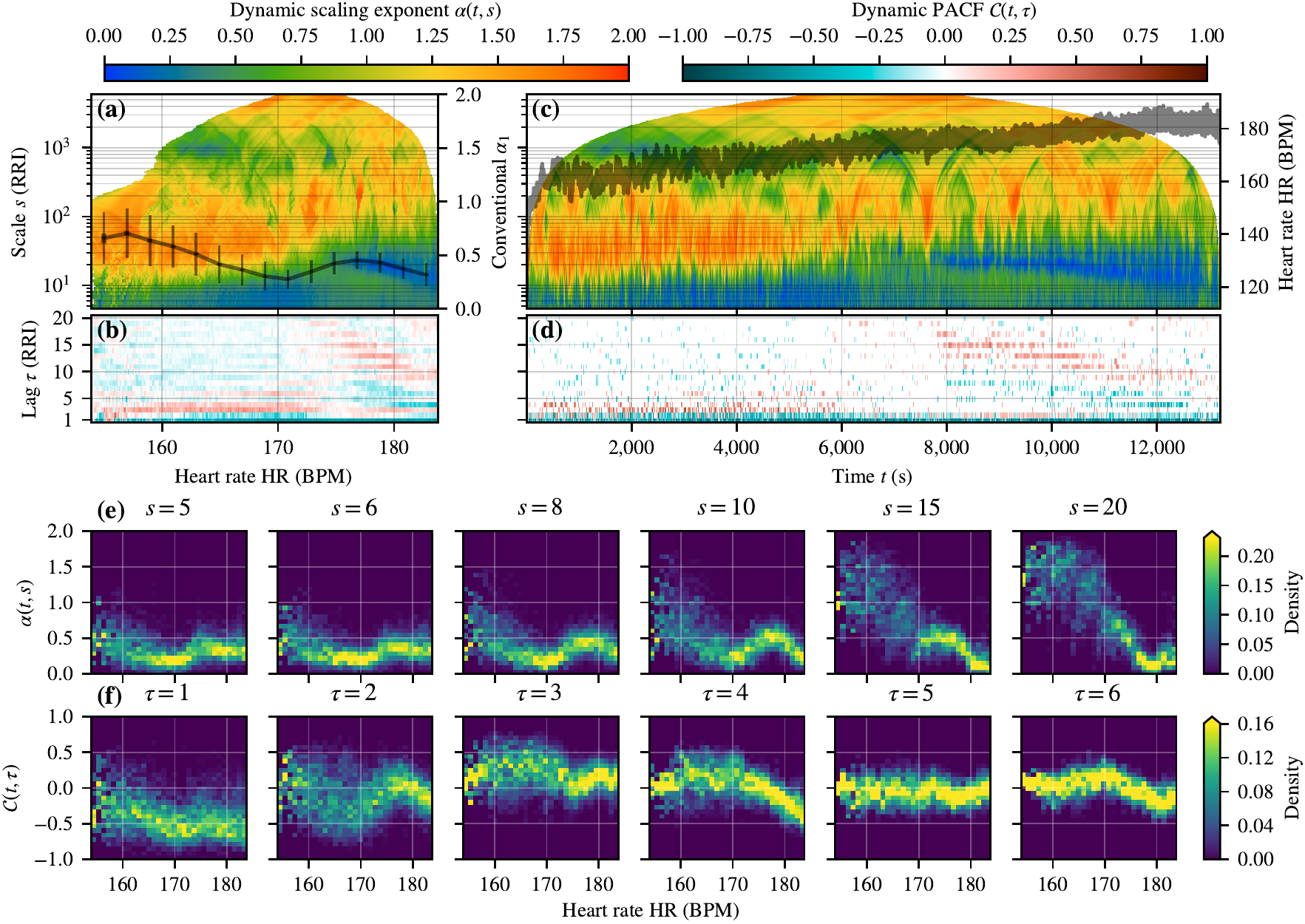}
    \caption{\label{fig:marathon example}
        Beat-to-beat (RR) interval correlations for the Marathon race of subject M1. Note that the upper-left and upper-right color bars refer to (a,c) and (b,d), respectively.
        {\bf (a)} Color-coded dynamic (DDFA-1) scaling exponent $\alpha(t,s)$ on different scales $s$ (y-axis) as a function of binned HR (x-axis). Here $\alpha(t,s)$ is averaged over those dynamic segments, whose average HR falls within \num{0.1} BPM wide bins. The values for empty bins are linearly interpolated if the gap does not exceed \num{0.5} BPM.
        The black solid line shows the mean together with the standard deviation (thin lines) and the the standard error of the mean (thick lines, barely visible) of the conventional short-scale (4--16 RRIs) scaling exponent $\alpha_1$. The exponent is computed in moving windows of 50 RRIs in HR bins of \num{2} BPM.
        {\bf (b)} Color-coded partial autocorrelation functions (DPACF-0) $\mathcal{C}(t,\tau)$ with different lags $\tau$ (y-axis) as a function of the binned HR.
        {\bf (c)} Similar to (a) but as a function of time during the marathon race. The instantaneous heart rate is overlaid on the data.
        {\bf (d)} Similar to (b) but as a function of time. The values that do not pass the non-zero significance test as described in the text are shown in white.
        {\bf (e)} Probability density histogram for $\alpha(t,s)$ for different scales $s$ as a function of the HR.
        {\bf (f)} Probability density histogram of the DPACF-0 for different lags $\tau$ as a function of the HR.
        The histograms in (e-f) consist of 31-by-31 bins, and the probability densities are separately normalized for each HR bin, so that they better depict the distributions as a function of the HR instead of measuring the prevalence of different HR regions. Furthermore, the color bar is capped at the 99.5th percentile to avoid outliers dominating the color scale.}
\end{figure*}

Figure~\ref{fig:marathon example} demonstrates our methods applied to a single marathon run (subject M1) of group M. The color-coded value of the scale-dependent exponent $\alpha(s)$ is shown in the first row as a function of the binned heart rate (HR) [Fig.~\ref{fig:marathon example}(a)] and also as a function of running time $t$ [Fig.~\ref{fig:marathon example}(c)].
Over the studied scales $s$ from 5 to 5000 heart beats, the scaling exponent $\alpha(s)$ exhibits complex behavior that could not be adequately described by the conventional division into short- and long-range scaling exponents. We consider the HR-dependent shift to anticorrelated RRIs at the shortest scales $s \lesssim$ 10--30 as the most interesting of our observations.
As the heart rate increases the anticorrelations extend to slightly longer scales until there is a qualitative change at approximately 175 BPM. The strongest anticorrelations shift from the shortest scales to roughly 20 beats, and gradually refill the shortest scales as the HR is increased even further. At larger scales $s \gtrsim 100$ the RRIs become mostly non-stationary ($\alpha > 1$, fractional-Brownian-motion-like behavior).
In contrast, a typical 24-hour RR-tachogram of a healthy subject at rest usually displays $1/f$ or pink noise on long time scales (or low frequencies $\lesssim 0.05$ Hz), corresponding to $\alpha=1$, and larger values for $\alpha$ at the shortest scales or higher frequencies \cite{Goldberger2002}.

The black curve in Fig.~\ref{fig:marathon example}(a) corresponds to the conventional DFA exponent $\alpha_1$ over the scales from 4 to 16 heart beats. It shows almost linear decrease to values around $1/2$ and below, when the DDFA exponent $\alpha(s)$ displays short-scale anticorrelated behavior. However, this simpler estimate is not sufficient for uniquely distinguishing the presence of anticorrelations from their shift to slightly longer scales.
To explore the scale dependence of the anticorrelations in more detail, we show in Fig.~\ref{fig:marathon example}(e) the probability density for the values of $\alpha(s)$ for six different scales $s$ from 5 to 20 heart beats as a function of the binned HR. On all six scales, the probability is maximum for $\alpha<1/2$ with a HR dependent modulation and an absence of anticorrelations on the two largest scales $s=15, 20$ for lower beat rates.

In order to estimate the relevant time scales of the physiological processes behind the observed anticorrelated beat intervals, we have also performed a DPACF analysis. The result is shown for lags between 1 and 20 heart beats in Figs.~\ref{fig:marathon example}(b) and \ref{fig:marathon example}(d). The PACF reveals direct anticorrelations (negative values) after a time lag of 1 and 2 beats, starting at low exercise intensities, and additional anticorrelations up to about 10 beats beyond HR of about 175 BPM, being consistent with the DDFA results.
The probability density of the DPACF values for lags between 1 and 6 beats, shown in Fig.~\ref{fig:marathon example}(f), confirm dominant direct anticorrelations on the shortest time scales of 1 to 2 beats, and 4 beats for high exercise intensities (here HR $\gtrsim 170$).

Subject M1 as the chosen example has the most prominent anticorrelations and particularly simple, almost linear, trend in the HR over the whole marathon. The {\em individual} DDFA and DPACF results, similarly to Fig.~\ref{fig:marathon example}, for all the subjects of group M are shown in Supplementary Fig.~\ref{fig:marathons overview}. The results share qualitative similarities across the subjects, as they all exhibit short-scale suppression of correlations and the appearance of anticorrelations as a function of the HR. However, some differences are also apparent, as only three subjects (M1, M3, and M7) show the shift of the anticorrelations to elevated scales at the highest exercise intensities. Some short-scale anticorrelations, particularly for subject M6 and to some extent for M4 and M5, also appear at elevated scales, but these happen at lower intensities and are likely different in origin. Regardless, additional research is required to determine the effect of individual strains relative to standard physiological thresholds on the results.

\begin{figure*}
    \includegraphics{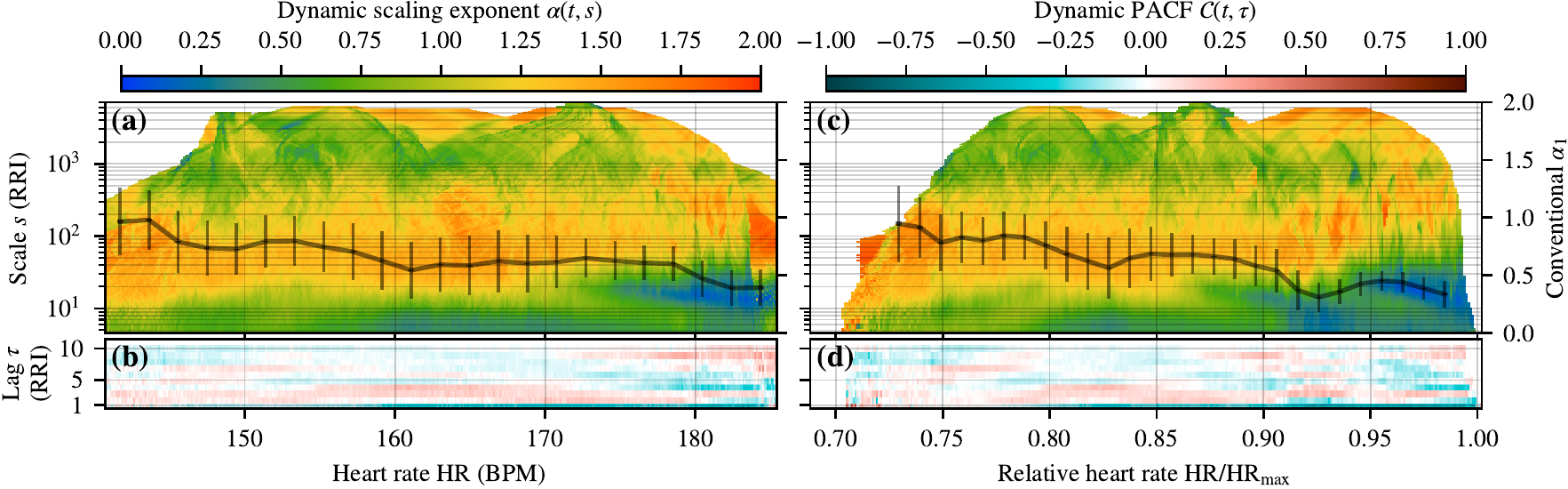}
    \caption{\label{fig:marathons aggregate}
        Aggregate beat-to-beat (RR) interval correlations as a function of heart rate for {\em all} subjects of group M. 
        {\bf (a)} Average values for $\alpha(t,s)$ for each scale $s$ (y-axis) and HR bin (x-axis). The solid line depicts the conventional short-range $\alpha_1$.
        {\bf (b)} Average values for $\mathcal{C}(t,\tau)$ for each lag $\tau$ (y-axis) and HR bin (x-axis). In (a-b), the data is processed as in Fig.~\ref{fig:marathon example}. 
        {\bf (c-d)} Similar to (a-b) but as a function of the relative HR. In (c-d), the
        data is processed as in Fig.~\ref{fig:marathon example}, but with the distinction that the relative HR bin width is \num{0.001}, the interpolation threshold is \num{0.005}, and the bin width for the conventional short-scale exponent $\alpha_1$ in (c) is \num{0.01}.}
\end{figure*}

To further study the consistency of the results between the different subjects, the aggregated DDFA (top) and DPACF (bottom) results for {\em all members of group M} are shown in Fig.~\ref{fig:marathons aggregate} as a function of both the absolute (left) and relative (right) HR. The most notable features are the high-intensity elevated-scale anticorrelations starting to appear at 87\% and 95\% relative HR, or at the absolute HR of 175 BPM (this congruence on the absolute scale is likely to be coincidental). In these ranges also the conventional DFA exponent $\alpha_1$ (black curve) drops slightly below $1/2$, but its limitations are apparent, as it is based on linear regression over the scales of 4 to 16 beats. The anticorrelations at lower intensities (approximately 155--175 BPM) appear to be more condensed on the relative scale (roughly 78--87\%, with a more concentrated maximum between 80--85\%), which is apparent both on the DDFA results and in the more pronounced dip of the short-scale exponent $\alpha_1$. There is also a band of short-scale suppressed correlations with a trend towards longer scales at even lower relative HR ($\approx$~72--79\%) that is practically indistinguishable on the absolute scale.
On the relative scale, the DPACF results also show a sharper transition into anticorrelated behavior at approximately 80\% HR for lag $\tau=1$, whereas on the absolute scale the transition is more gradual.
These lag $\tau=1$ anticorrelations appear consistently for all the subjects and become stronger with increasing HR, and also appear at longer lags at the regions where the DDFA anticorrelations shift to larger scales. Naturally, the aggregated results should be interpreted with care as they represent an average result over all the samples. Secondly, there is uncertainty in the maximum HR values of the subjects. Nevertheless, the results suggest that neither the absolute nor the relative scale is universal for different individuals.

\subsection*{Freeform Training Runs}

\begin{figure*}
    \includegraphics{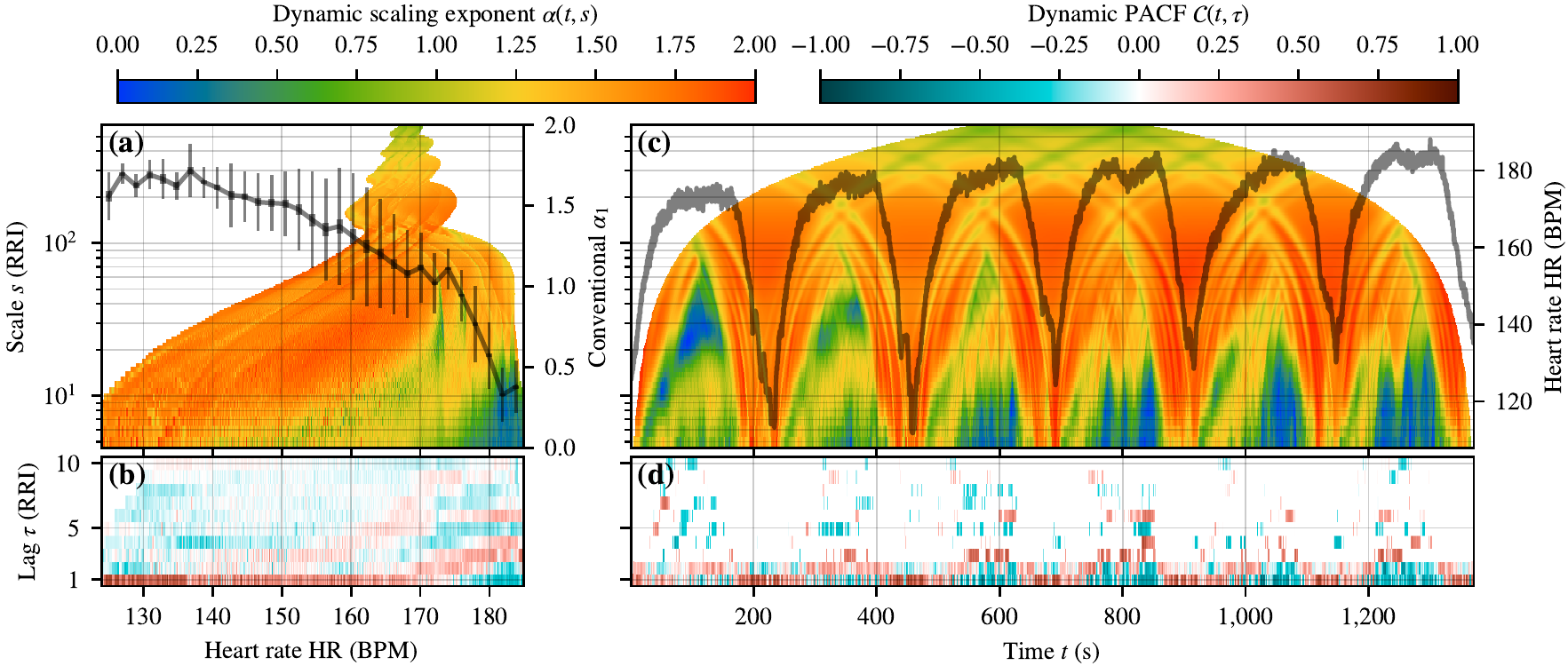}
    \caption{\label{fig:mit example}Beat-to-beat RR interval correlations for {\em one} interval exercise from group T. {\bf (a)} Dynamic scaling exponents (DDFA-1) $\alpha(t,s)$ (colors) and the conventional short-range $\alpha_1$ (solid line) as a function of the binned HR. {\bf (b)} DPACF-0 correlations $\mathcal{C}(t,\tau)$ as a function of the binned HR. {\bf (c-d)} As in (a-b) but as a function of time, and in (c) the HR value is overlaid on the data. For details on the data processing, see the caption of Fig.~\ref{fig:marathon example}.}
\end{figure*}

In order to study the correlations of RRIs over a wide range of exercise durations and intensities, we perform the same analysis for subjects in group T. It is instructive to consider first a single exercise of one subject which is shown in Fig.~\ref{fig:mit example}. It consists of six intervals of high-intensity running, each interval lasting about 160 seconds with the subsequent intervals reaching higher and higher intensities.
As a function of exercise time $t$ the DDFA exponent $\alpha(s)$ [Fig.~\ref{fig:mit example}(c)] and the PACF [Fig.~\ref{fig:mit example}(d)] consistently reveal strong anticorrelations of RR intervals that develop rapidly after the start of the intense interval. The shortest-scale anticorrelations span to longer and longer scales with increasing HR in the latter intervals. The earlier lower-intensity intervals exhibit anticorrelations at elevated scales, separated by a band of correlations from the shortest scale anticorrelations. This behavior was already suggested by some of the marathon data (M4, M5, and M6), and in a following analysis we will relate these to a distinct band of anticorrelations appearing at moderate exercise intensity.
At rest between the intervals the anticorrelations rapidly vanish.
The DPACF shows strong lag $\tau=1,2$ anticorrelations, whose magnitudes are in accordance with the short-scale DDFA anticorrelations as observed in group M.
The existence of patches of anticorrelations over time lags up to 10 beats is also consistently observed with the elevated-scale DDFA anticorrelations.
As a function of HR, the anticorrelated behavior develops rapidly after an intensity threshold ($\approx$~175 BPM) [see Fig.~\ref{fig:mit example}(a)]. The elevated-scale anticorrelations are visible as a spike of suppressed correlations at approximately 172~ BPM with a weak tail towards short scales and lower intensities. This latter phenomenon is more visible in the DPACF data [Fig.~\ref{fig:mit example}(b)] as a weak band of anticorrelations surrounded by correlated bands at shorter and longer lags.

\begin{figure*}
    \includegraphics{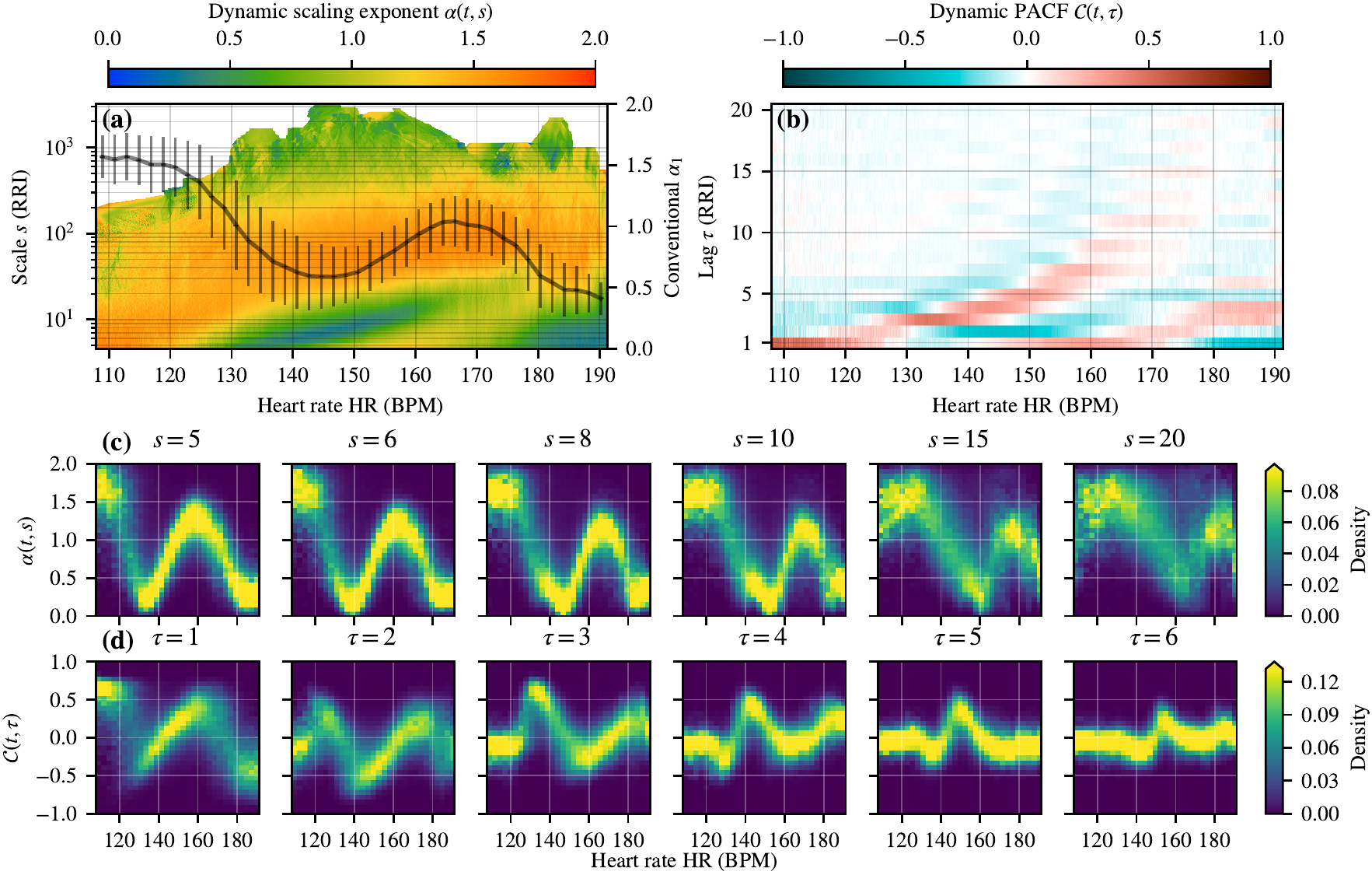}
    \caption{\label{fig:hrv7 aggregate}Aggregate beat-to-beat (RR) 
        interval correlations for \emph{all} the exercises of one subject (T07) in group T.
        {\bf (a)} Average values for $\alpha(t,s)$ for each scale $s$ (y-axis) and binned HR (x-axis). The solid line shows the conventional short-range $\alpha_1$.
        {\bf (b)} Average values for $\mathcal{C}(t,\tau)$ for each lag $\tau$ (y-axis) and binned HR (x-axis).
        {\bf (c)} Probability density histogram for $\alpha(t,s)$ for different scales $s$ as a function of the HR.
        {\bf (d)} Probability density histogram for $\mathcal{C}(t,\tau)$ for different lags $\tau$ as a function of the HR.
        Note that the probability densities are separately normalized for each HR bin.
        For details on data processing, please see the caption of Fig.~\ref{fig:marathon example}.}
\end{figure*}

Next, we study the typical behavior of RRI correlations when averaged over many running exercises of different intensity and duration, but for the same subject to avoid effects due to individual variability. The corresponding aggregated results from DDFA and DPACF analysis for the subject T07 are shown in Fig.~\ref{fig:hrv7 aggregate}, representing a total running distance of 1889 km. For this large data set, we obtain good statistics for the aggregated data and expect them to provide a reliable representation of the typical RRI correlations as function of the exercise intensity. Indeed, both DDFA exponent $\alpha(s)$ and DPACF clearly show two distinct bands of anticorrelated RRIs [see Fig.~\ref{fig:hrv7 aggregate}(a) and (b)].
The first band at moderate exercise intensity ($\approx$~125--170 BPM) displays a distinct, approximately exponential, trend in the DDFA anticorrelations shifting to longer scales as a function of the HR.
It is plausible that the elevated scale anticorrelations appearing at lower intensities in Fig.~\ref{fig:mit example} and for M4, M5, and M6 originate from this band of anticorrelations.
The corresponding band in the DPACF results is split by a band of strong positive correlations.
The latter band of anticorrelations at high exercise intensities ($\gtrsim 175$ BPM) does not show a clear trend as a function of the HR, although there is tendency towards spanning to longer scales with increasing intensity. Notably, the anticorrelations remain present even at the shortest scales and lag.
This is in contrast to some of the marathon data, where the highest-intensity anticorrelations appear at elevated scales. This could be due to the different nature of the exercises, as in the marathon races these anticorrelations appear after prolonged exercise at high intensity, and changes in, e.g., body temperature or electrolyte balance may influence the results.
On the other hand, in the discussion section we make an argument that this could be due to interactions with the stride frequency.
The conventional $\alpha_1$ indicates the suppression of correlations that is consistent with the DDFA anticorrelations when taking into account its limitation to the scales of 4--16 beats. The $\alpha_1$ is clearly insufficient to capture anticorrelated behavior concentrated on thin bands of scales.
Figures~\ref{fig:hrv7 aggregate}(c) and \ref{fig:hrv7 aggregate}(d) show the probability density plots of $\alpha(s)$ and PACF values for different scales $s$ and lags $\tau$, respectively.
The existence of two regions with anticorrelated  RRIs is clearly visible. They are separated by a region with positive correlations (or $\alpha>1/2$).

The aggregated data for all the subjects of group T are shown in Supplementary Fig.~\ref{fig:mit bpm aggregate}. Most subjects show common qualitative similarities in the form of two anticorrelated bands as described for T07. However, for some subjects the split into the two anticorrelated regions is not that clear; particularly there is the lack of correlated shortest scale behavior separating these two regions.
In the absence of the correlated bands the behavior is remarkably simple; the higher the intensity, the more prominent the anticorrelations are in both magnitude and scales covered.
In addition to individual intrinsic cardiac variability, a possible explanation could be different training practices and external conditions, as for example T05 shows behavior that is most similar to the marathon data.
Another explanation could be highly regular running motion, which could promote correlations induced by, e.g., muscle contractions and blood pressure variations, which is an argument set forth in the next section.
It is also worth noting that T11 reported problems with the chest strap, and as a result his data has unusually high amount of missed beats (up to 50\%). Despite of this, two regions of suppressed correlations are present that are consistent with the other subjects, highlighting the robustness of the methodology.

We assessed the suitability of higher order detrending for our analysis and decided to employ DDFA-1 due to the following reasons: (i) The qualitative behavior remains the same at the shortest scales, which is the most interesting region for dynamic exercise intensity analysis, (ii) the short-scale bias in DFA is larger and crossover scales are shifted with higher order methods, (iii) higher orders of DDFA appear to require longer dynamic segments for similar statistical accuracy and have increased computational cost.

Finally, we point out that it is important to check the reliability of our DDFA and DPACF methods with respect to trends. Hence, we have filtered the data of subject T07 according to the condition that the standard deviation of the HR within the dynamic segments is smaller than the values for certain quantiles. We find that the observation of the bands with anticorrelations is robust and independent of the choice of the quantile filter. In fact, the anticorrelations appear stronger when limiting to dynamic segments with less HR variation, as the averaging is not performed over segments with transient changes in the HR that could lead to spurious correlations. The exact results of this analysis for six different choices of quantiles are shown in Supplementary Fig.~\ref{fig:results trend consistency}.

\section*{Discussion\label{sec:discussion}}

It is important to understand the physiological mechanism causing the observed anticorrelations.
Due to the lack of time series for other physiological variables, we can present below only simple arguments that we consider to be potentially relevant for explaining the observed dynamic correlations. 
First, we point out that there are three physiologically relevant time scales that fall into the range over which the anticorrelations occur: (i) the stride frequency which is typically around 85 strides per leg and per minute \cite{Lieberman+2015}, (ii) the respiration cycle which is typically three to five heart beats long, and (iii) arterial blood pressure fluctuations, i.e., the so-called Mayer waves, which result from an oscillation of sympathetic vasomotor tone and is of the order of ten seconds \cite{Julien2006}.

All three processes are cyclic and hence can induce periodic modulations to the heart rate through hemodynamics. Such periodicities could result in anticorrelated RRIs when observed at scales similar to the period measured in heart beats.
Furthermore, the overall heart rate variability is reduced under exercise due to withdrawal of cardiac vagal tone and parasympathetic control\cite{Karasik2002,Hautala2003,Gronwald2019}, i.e., the local short-term  RRIs are more regular without $1/f$- or Brownian-like diffusion for extended periods of time.
Therefore, subtler patterns should become more discernible, as they are not masked by the complex fluctuations of a healthy heart under resting conditions.
Similarly, the relative magnitudes of the modulating signals could affect the scale-dependence of the anticorrelations.
If some of the effects is much stronger, it will mask higher frequency periodicities as they will appear correlated when superimposed on the stronger lower frequency oscillations.
Additionally, when a periodic signal is sampled at discrete intervals, the result is a new signal whose period depends on the sampling frequency. This effect is manifested, e.g., when the influence of the blood pressure variations due to the stride frequency is sampled at each heart beat.

This latter phenomenon could explain some of the qualitative differences in the dynamic correlations between the subjects.
For some subjects (particularly T08, but also T02, T03, T10, and T12) the RRIs show clearly defined behavior under exercise, becoming short-scale anticorrelated at moderate intensity, with the magnitude and the scale of the anticorrelations increasing in conjunction with intensity. In contrast other subjects exhibit more complex RRI-correlations where the simple anticorrelations are interrupted by bands of decreased or altered correlations at shorter scales (please see Supplementary Figs.~\ref{fig:mit bpm aggregate} and \ref{fig:marathons overview} for the individual RRI-correlation plots as a function of the HR).
These more correlated bands appear at heart rates corresponding to sampling frequencies where typical stride frequencies would look correlated at the shortest scales.
If these bands arise from the stride frequency, that could also explain the better congruence on the absolute HR scale in Fig.~\ref{fig:marathons aggregate}, as there is generally less variance in the stride frequencies than in the maximum heart rates.

These considerations would imply that the (anti)correlations arise from underlying universal cardiolocomotor mechanisms, but detailed response to exercise may depend on individual physiology, biomechanics of running and training status \cite{DiMichelle2012}.
Furthermore, the onset of the anticorrelations, their strength, and scales of appearance show individual variability.
Studying the relationship of these variables to standardized thresholds and markers in exercise physiology could allow utilizing the dynamic correlations for monitoring the exercise intensity in real-time without the knowledge of parameters such as the maximal oxygen uptake (VO2max) or maximum heart rate.
We are aware of the possibility that the universal emergence of anticorrelations at elevated heart rates is most likely affected by other physiological factors beyond the ones discussed here.
Clearly, further research is required, but the approach herein provides a promising avenue forward.

\section*{Conclusions\label{sec:conclusions}}

Our main result is the discovery of multiscale anticorrelations in RR intervals during running exercises under real-world conditions.
The anticorrelations have a dynamical structure that depends on the exercise intensity as measured by the heart rate.
The characteristics of the dynamical structure are revealed by our methodology, in particular the dynamic detrended fluctuation analysis and dynamic partial autocorrelation functions, which we anticipate becoming useful tools in data analysis across various disciplines. While we have demonstrated the capability to study the dynamical RRI correlations during varying real-world circumstances, a more systematic evaluation of the methodology is required to control for exercise conditions.

The observed anticorrelations appear on short scales (a few beats) at low to moderate exercise intensities. As the intensity is increased, the anticorrelations increase in magnitude and span to longer scales (up to 20--30 beats).
This simplified picture is complicated by correlations arising potentially from interactions with the regular running motion when the stride frequency is appropriately proportional to the heart beat. These correlations mask the anticorrelated behavior on bands of increasing or decreasing scales at moderate and high exercise intensities, respectively.
At rest, e.g., between running intervals, the anticorrelations rapidly vanish, and appear immediately when the intensity is increased again.
These changes happen before the HR saturates at the level necessary to maintain the ongoing exercise intensity.
Hence, our findings allude the possibility of quantifying the relative exercise intensity by measuring the dynamic correlation exponent $\alpha(t,s)$ in real time during exercise.

This report of our initial findings serves as a prelude for highlighting the potential of the dynamic correlation analysis so that further advances could be pursued.
It is highly desirable to develop a theoretical model for the complex dynamics of the cardiovascular feedback loops during high-intensity exercise load that can explain the observed time scales for the anticorrelated RR intervals. Clearly, a more systematic study with subjects performing specific exercise protocol should be performed to verify our observations. Besides, a thorough validation and calibration of our results with data collected during running exercise in a physiology laboratory is a natural next step for our study to relate the changes in the dynamic correlations to standard exercise physiology models. The inclusion of accelerometer data, from which the stride frequencies could be derived, would facilitate the verification of the running modalities as a possible cause for the bands of correlations present for some subjects.

Such controlled and systematic studies are not only necessary to elucidate the speculative nature of the results herein, but they are further motivated by potentially enabling the application of this methodology in exercise physiology.
We expect that the reported RR interval correlations are suitable to represent a dynamical ``fingerprint'' of the exercise-induced cardiovascular load. Hence, our methodology -- which could be integrated with the present devices on the market -- has a potential to become a new tool in real-time exercise monitoring without previous knowledge of maximal thresholds such as the maximum hearth rate and lactate or ventilatory thresholds.

\section*{Materials and Methods}{\label{sec:materials}}

\subsection*{Heart Rate Data during Exercise{\label{sec:data}}}

We study real-time correlations of RRIs during exercises of various intensities.
All heart rate data for this study have been collected during regular running training and racing under real-world conditions, i.e., outside the laboratory. Two groups of data were used for our theoretical analysis. The first group of data was recorded by human volunteers during their regular running training with freely chosen intensity and volume (group T). The study period was at least 4 weeks, and some subjects provided data over a longer period of time. We obtained institutional approval and informed consent
(COUHES exemption for the employed protocol has been granted under protocol no.~1711132002).
The research was performed in accordance with the rules and regulations set by the participating universities.
This group involved 12 volunteers (5 female, 7 male, with an age span from 27 to 65 years). Their performances span a wide range from top national level to recreational runners: the personal bests in 10 km range from 29 min 31 sec to 44 min 57 sec, in marathon from 2 hours 43 min 20 sec to 4 hours 26 min 3 sec. 

During exercises, heart rate (HR), RR intervals (RRI), running velocity and distance were recorded using a Garmin heart rate monitor HRM4-Run and a GPS watch (Forerunner 935, Garmin Inc., Olathe, KS, USA). A previous study has investigated and validated the accuracy of this HRM \cite{Cassirame2017}. The data were recorded by the GPS watch in the Flexible and Interoperable Data Transfer (FIT) format \cite{FIT_web} and subsequently uploaded by the subjects to a web service that we had launched for this study. The total number of exercise files analyzed per subject (samples) varied between 18 and 261, with total covered distances from 150 to 1889 km. 

The second group of data was obtained by selecting randomly the marathon races of 7 subjects from data uploaded to the Polar Flow web service\cite{Polar_web} (group M). Within registration to Polar Flow, the subjects have given their consent for the use of their data for research purposes. The metadata were provided by the users of this web service (all male, with an age span from 28 to 53 years, and marathon finishing times between 3 hours 30 min and 4 hours 17 min). HR and RRIs were recorded for this group of subjects with a Polar heart rate monitor H10 and a Pro Strap (Polar Electro Oy, Kempele, Finland). Recently, the RR signal quality of this HRM has been shown to be excellent from low- to high-intensity activities in comparison to a ECG Holter device \cite{Gilgen2019}.  In both groups T and M, the subjects provided their maximum and resting heart rates. Summaries of all the metadata for the two groups are shown in Supplementary Table~\ref{tab_data}.

As ECG data is not available, we do not attempt to remove ectopic beats or other artifacts based on physiological criteria. Therefore we merely remove technical artifacts, such as missed beats, that can be isolated with reasonable certainty. The details for this data preprocessing are provided in Supplementary Appendix~\ref{sec:data preprocessing}.

\subsection*{Conventional Methods\label{sec:conventional methods}}
For comparison we apply ordinary detrended fluctuation analysis \cite{Peng1994,Kantelhardt2001} to the RRI time series.
By computing the root-mean-squared fluctuations $F(s)$ around local trends at multiple scales $s$, the method assesses power law scaling relations $F(s) \propto s^{\alpha}$ characterized by the scaling exponent $\alpha$. In the context of HRV, typically two exponents are determined, for short- ($\alpha_1$) and long-scale ($\alpha_2$) correlations, respectively \cite{Peng1995,Shaffer2017}.
We extract the conventional short-scale (4--16 RRIs) scaling exponents $\alpha_1$ \cite{Peng1995} in segments consisting of 50 RRIs.
We compute the fluctuation functions in maximally overlapping windows for enhanced statistical properties \cite{Kiyono2016}. A summary of the DFA method is provided in Supplementary Appendix~\ref{sec:dfa}.
We also provide a helpful summary of partial autocorrelation functions in Supplementary Appendix~\ref{sec:autocorrelation} before introducing their dynamic counterparts here.

\subsection*{Dynamic Segmentation\label{sec:dynamic segmentation}}
The dynamic behavior of the time series can be studied by performing the analysis in moving temporal segments. However, to guarantee sufficient statistical accuracy, the length of these segments is dictated by the largest scale $s$ (DFA) or the lag $\tau$ (PACF), resulting in diminished temporal resolution for small scales. Therefore, we propose a dynamic segmentation procedure, where the segment length is varied as a function of the scale or the lag:
\begin{enumerate}
    \item Choose a function for determining the segment lengths $\ell(s)$ as a function of the scale $s$. Here we adopt a simple linear relationship $\ell(s) = \dlf s$ where $\dlf$ is a constant. Smaller values increase the temporal resolution but also the statistical noise. The dynamic length factor $\dlf$ itself may also be varied for different scales.
    \item For each scale divide the time series into segments of length $\ell(s)$. The segments themselves may be overlapping if desired for smoother results. Identify the segments $\mathcal{S}_{s,t}$ by their temporal indices $t$, which may be, e.g., the mean time within the segment or any other suitable quantity.
\end{enumerate}

\subsubsection*{Dynamic Detrended Fluctuation Analysis (DDFA) \label{sec:dynamic dfa}}

The dynamic segmentation together with the maximally overlapping windows in the DFA scheme enables the following procedure for dynamic DFA (DDFA):
\begin{enumerate}
    \item Perform the dynamic segmentation for each scale $s$. The value of $a=5$ was found to be an acceptable value for the dynamic length factor, which is employed in all of our DDFA calculations.
    \item Utilizing overlapping windows, compute the fluctuation function in each segment $\mathcal{S}_{s,t}$ at scales $\{s-1, s, s+1\}$. Denote the logarithmic fluctuation function at these scales by $\Fsm, \Fs$ and $\Fsp$, respectively.
    \item In each segment, compute the dynamic scaling exponent $\alpha(t,s)$ by the finite difference approximation \cite{Fornberg1988}
    \begin{align}
    \FiniteDifferenceApproximation \text{,}
    \end{align}
    where $\hs = \log(s) - {\log(s - 1)}$ and $\hd = {\log(s + 1)} - \log(s)$ are the logarithmic backward and forward differences. Fluctuation functions computed with maximally overlapping windows are empirically found to be smooth enough to permit the direct application of the finite difference scheme.
\end{enumerate}

The performance of the method is numerically validated by applying it to simulated time series with known properties. Supplementary Appendix~\ref{sec:app_a} explains the details for analytically obtaining the theoretically expected scale-dependency of DFA scaling exponents for different processes. In Supplementary Appendix~\ref{sec:validation of the methods} these theoretical results are utilized for confirming the acceptable performance of the DDFA method.

\subsubsection*{Dynamic Partial Autocorrelation Function (DPACF)\label{sec:dynamic_pacf}}

In order to obtain a {\it local} estimate of the partial autocorrelation function $\cC(\tau)$  we compute it using an approach similar to that of the DDFA algorithm. The steps of this approach can be summarized as follows:
\begin{enumerate}
    \item Perform dynamic segmentation for each lag $\tau$. The value of $a=10$ was found to be an acceptable value for the dynamic length factor, which is utilized in all of our DPACF calculations.
    \item In each segment $\mathcal{S}_{\tau,t}$, perform polynomial detrending of order $m$.
    \item For each segment, compute $\cC(\tau)$ by, for example, solving the Yule-Walkers equations with the Levinson-Durbin recursive scheme \cite{Durbin1960}. Choose each time for the maximum lag the parameter for which we are estimating the partial autocorrelation function. Denote this dynamic PACF by $\cC(t,\tau)$.
\end{enumerate}

Resorting to the central limit theorem, it is a known result that the partial autocorrelation function is approximately non-zero at 5\% significance level if $\abs{\cC(t,\tau)}< 1.96/\sqrt{\ell(\tau)}$.
The evaluation of this significance band is statistically valid only if $\ell(\tau)\gtrsim 30$ and therefore if $\tau>3$ for $a=10$.

Notice that in DPACF the detrending is applied to the original time series in contrast to the integrated series in DDFA. Therefore, the results would be expected to be qualitatively similar when the DPACF detrending order is one smaller than the DDFA detrending order $n$. This explanation is complicated by, e.g., the removal of linear correlations in PACF. However, the relationship ${m \approx n - 1}$ is supported by empirical observations.

While both DDFA and DPACF measure dynamic correlations, it is important to realize the qualitative difference between them. DDFA describes the \emph{collective} behavior of all beats over the scale $s$, whereas DPACF considers the average behavior of \emph{individual} beats separated by the lag $\tau$ (with the linear dependence from the preceding lags removed) 

\begin{acknowledgments}
We acknowledge insightful discussions with Ary Goldberger. We also thank Jyrki Schroderus, Laura Karavirta, Jussi Peltonen and Arto Hautala for all the support and suggestions. The authors acknowledge Polar Electro Oy (Oulu, Finland) for providing exercise data and CSC -- IT Center for Science, Finland, for computational resources. Additional support was provided by an EMERGENCE grant of CNRS, INP, and the ICoME2 Labex (ANR-11-LABX-0053) and the A*MIDEX projects (ANR-11-IDEX-0001-02) funded by the French program ``Investissements d'Avenir'' managed by ANR.
 \end{acknowledgments}

\appendix
\setcounter{figure}{0}
\setcounter{table}{0}
\renewcommand{\thetable}{S\arabic{table}}
\renewcommand{\thefigure}{S\arabic{figure}}
\renewcommand{\figurename}{Supplementary Figure}
\renewcommand{\tablename}{Supplementary Table}

\section{Dataset details and preprocessing\label{sec:data preprocessing}}
We study two datasets of running exercises and races under real-world conditions. These consist of voluntary running exercises at freely chosen intensities and schedules (group T) and marathon races (group M).
Metadata for these datasets are presented in Supplementary Table \ref{tab_data}.
\begin{table*}
    \caption{\label{tab_data}Summary of the metadata for the two groups of subjects: Training runs of various durations and intensities (T) and marathon races (M). For group T shown are the age, gender, resting heart rate HR$_\text{rest}$ and maximum heart rate HR$_\text{max}$ (as reported by the subject), the personal best (PB) time for 10 km and marathon (time in hh:mm:ss) within 3 years before this study, the number of exercise samples analyzed, the average heart rate HR$_\text{avg}$ of all samples, and the total distance and duration of all samples. For group M all subjects were male, and shown are their age,  resting heart rate HR$_\text{rest}$ and maximum heart rate HR$_\text{max}$ (as reported by the subject), marathon finishing time, and average heart rate HR$_\text{avg}$ during the marathon.}
    \begin{tabular}{ccccccccccc}
        \multicolumn{11}{c}{Group T}\\
        \toprule
        subject & age [y] & gender & HR$_\text{rest}$ & HR$_\text{max}$  & PB 10 km &  PB Marath. &samples & HR$_\text{avg}$ & distance [km] & \multicolumn{1}{c}{duration [h]} \\
        \midrule
        T01 & 48 & m & 40 & 192 & 00:36:49 & 02:50:49 &  186 & 151 & 1527 & 121.5\\
        T02 & 27 & m & 40 & 193 & 00:29:31 & -- &  54&  130& 576 & 43.5\\
        T03 & 29 & f & 43 & 200 & 00:36:50 & 02:47:56 &  20&  155& 170& 13.7\\
        T04 & 29 & f & 50 & 205 & 00:42:30 & -- &  103&  167& 1076& 92.8\\
        T05 & 39 & f & -- & 200 & 00:44:57 & 03:40:09 &  15&  176& 150& 11.7\\
        T06 & -- & m & 40 & 192 & 00:35:22 & 02:43:25 & 26 & 153 & 666 & 49.3 \\
        T07 & 33 & m & 42 & 214 & 00:33:56 & 02:43:20 & 261 & 154 & 1889 &138.7\\
        T08 & 37 & f & 43 & 200 & -- & 04:26:03 &  20&  149& 209& 22.1\\
        T09 & 27 & m & 48 & 195 & 00:31:32 & -- &  21&  143& 199& 16.7\\
        T10 & 37 & f & 45 & 179 & -- & -- &  18&  154& 215& 18.1\\
        T11 & 65 & m & 47 & 170 & 00:43:10 & 03:13:29 & 26 & 134 & 287& 28.2\\
        T12 & 47 & m & 48 & 182 & -- & 03:09:49 & 53 & 129 & 1133& 95.4\\
        \bottomrule
    \end{tabular}
    \begin{tabular}{cccccc}
        \multicolumn{6}{c}{Group M}\\
        \toprule
        subject & age [y] & HR$_\text{rest}$ & HR$_\text{max}$ & Finishing time [hh:mm:ss] & HR$_\text{avg}$ \\
        \midrule
        M1 & 53  & 56& 185& 03:40:09 & 172\\
        M2 & 50 & 46& 174& 03:36:47 & 149\\
        M3 & 28 & 58& 198& 04:17:04 & 171\\
        M4 & 34 & 50& 195& 04:12:36 & 155\\
        M5 & 43 & 55& 200& 03:49:03 & 162\\
        M6 & 39 & 55& 194& 04:19:36 & 155\\
        M7 & 50 & 50& 200& 03:30:33 & 174\\
        \bottomrule
    \end{tabular}
\end{table*}

Artifacts in the data are removed prior to the analysis by utilizing the following scheme:
\begin{enumerate}
    \item RR intervals below or above the threshold values $\RRmin$ and $\RRmax$ are removed.
    \item Compute the local median of the RR intervals $\RRmed(t)$ in a moving window of length $\lengthmed$.
    \item Remove RR intervals that fall outside the range $(1 \pm \thresholdmed)\RRmed(t)$, where the threshold $\thresholdmed$ is a constant.
\end{enumerate}
The values for the filtering parameters are listed in Supplementary Table \ref{tb:filtering parameters}. Acceptable performance of the filter is manually inspected for all the samples in group M and a representative subset of 17 samples from the group T. The filter is particularly adept at removing too long intervals arising from missed beats, which is the most common error in exercise data \cite{Giles2018}. 
As ECG data is not available, we do not attempt to filter the data based on physiological criteria, and merely remove technical artifacts that can be isolated with reasonable certainty.
\begin{table}
    \caption{\label{tb:filtering parameters}Data filtering parameters.}
    \begin{tabular}{ccccc}
        \toprule
        Group & $\RRmin$ (ms) & $\RRmax$ (ms) & $\lengthmed$ (beats) & $\thresholdmed$ \\
        \midrule
        M & 250 & 600 & 15 & \num{0.026} \\
        T & 250 & 1000 & 11 & \num{0.03} \\
        \bottomrule
    \end{tabular}
\end{table}

\section{Detrended Fluctuation Analysis (DFA) \label{sec:dfa}}

Ever since its introduction in the study of correlations in DNA sequences \cite{Peng1994}, DFA for time series has been widely employed across multiple disciplines such as physics \cite{Santhanam2006,Kotimaki2013}, medicine \cite{Peng1995,Goldberger2002,Kim2019}, finance \cite{Vandewalle1997}, and even music \cite{Hennig2011,Rasanen2015}. The DFA method has been extensively studied \cite{Kantelhardt2001,Hu2001,Chen2002,Heneghan2000,Kiyono2015,Holl2015,Lovsletten2017,Holl2016,Holl2019} and it has been expanded to account for effects such as multifractality \cite{Kantelhardt2002} and cross-correlations \cite{Podobnik2008}.

We briefly summarize the conventional DFA algorithm which has been developed to detect correlations in non-stationary time series \cite{Peng1994,Peng1995}. First, for a time series $X(j)$ of length $N$ a cumulative summation is performed,
\begin{align}
Y(k) &= \sum_{j=1}^{k} \left( X(j) - \mean{X} \right) \, ,
\label{eq:integrated time series}
\end{align}
where the mean $\mean{X}$ of the time series is subtracted, but that is not strictly necessary for DFA \cite{Kantelhardt2001}.
Conventionally, the integrated time series of \eqref{eq:integrated time series} is divided into non-overlapping windows of length $s$.
In each window $w$, a local trend is determined as the least-squares fit of a low order polynomial $p_{s,w}(k)$ to the data. (The method is denoted by DFA-$n$ if the degree of the detrending polynomial is $n$ \cite{Kantelhardt2001}.) The fluctuations are measured as the variance from the local trend $p_{s,w}(k)$ in each window: $F_{s,w}^2=\frac{1}{s}\sum_{k \in w}\left(Y(k) - p_{s,w}(k)\right)^2$. These squared fluctuations are averaged over the windows to yield the fluctuation function
\begin{equation}
\label{eq:dfa fluctuation function}
F(s) = {\mean{F_{s,w}^2}}^{1/2} \text{.}
\end{equation}
Allowing the windows to overlap enhances the statistical properties of this estimate \cite{Kiyono2016}.
When this procedure is repeated for different window sizes, or scales $s$, a power-law increase of the fluctuations with the window size may be observed, i.e., $F(s) \sim s^\alpha$. Here $\alpha$ is a scaling exponent that can be considered as a generalization of the Hurst exponent $H$ \footnote{Originally it was the exponent in Hurst's R/S analysis \cite{Hurst1956}, and Mandelbrot and van Ness used it as a parameter for defining fGn and fBm \cite{Mandelbrot1968}. For these processes it can be related to the (asymptotic) DFA scaling exponent (fGn: $\alpha=H$, fBm: $\alpha=H+1$). In the $0 < \alpha < 1$ range the relationship (between R/S and DFA exponents) is approximate, and for fGn/fBm holds asymptotically for large scales. For $\alpha > 1$ the interpretation $\alpha=H+1$ is essentially due to the definition of fBm. It also could be argued to hold, at least approximately, for processes that are defined as the cumulative sum of another process, similarly as the (discrete) fGn/fBm.}. See Supplementary Appendix~\ref{sec:validation of the methods} below for more details.

However, experimental time series rarely exhibit exact scaling over several scales.
Many previous studies have focused on finding a robust determination of the scaling regimes \cite{Ge2013,Habib2017}, or on extracting a {\em spectra} of scaling exponents $\alpha(s)$ \cite{Viswanathan1997,Echeverria2003,Castiglioni2009,Xia2013,Molkkari2018}. These methods are based on the notion that the spectra may be defined as the {\em local slope} of the logarithmic fluctuation function,
\begin{align}
\alpha(s) &= \frac{\dd{\left[\log F(s)\right]}}{\dd{\left[\log s\right]}} \text{.}
\label{eq:local alpha}
\end{align}
In the context of HRV, these methods generalize and expand the conventional division into short (4--16 beats) and long-range (16--64 beats) scaling exponents. In practice, the behavior may also change over time, either due to external influences, or the process itself may comprise several distinct intrinsic modes. This paper develops a methodology that takes these temporal variations into account in a consistent manner.

Depending on the value of the exponent $\alpha$, different degrees of correlations of the time series or its increments can be identified. The meaning of the different ranges for $\alpha$ are summarized in Supplementary Table~\ref{tb:dfa scaling exponent}. 
\begin{table}
    \caption{\label{tb:dfa scaling exponent}Meaning of values of the DFA scaling exponent $\alpha$.
        The qualitative interpretation remains the same for even higher exponents that become discernible with higher-order DFA: For each integral interval the lower and upper halves correspond to originally anticorrelated or correlated increments, respectively.}
    \begin{tabular}{ccc}
        \toprule
        Scaling exponent & Interpretation & Stationarity \\
        \midrule
        $0 < \alpha < \nicefrac{1}{2}$ & anti-correlated & \multirow{4}{*}{stationary} \\
        $\alpha = \nicefrac{1}{2}$ & white noise \\
        $\nicefrac{1}{2} < \alpha < 1$ & correlated \\
        $\alpha = 1$ & $1/f$ (pink) noise & \\ \hline
        $1 < \alpha < 1\nicefrac{1}{2}$ & anti-correlated increments & \multirow{3}{2.5cm}{\centering non-stationary, stationary increments} \\
        $\alpha = 1\nicefrac{1}{2}$ & Brownian noise \\
        $1\nicefrac{1}{2} < \alpha < 2$ & correlated increments \\
        \bottomrule
    \end{tabular}
\end{table}

The scaling exponent $\alpha$ is related to other scaling exponents in time series analysis. Scale invariance is also observed in the Fourier domain as a function of the frequency $f$ with a power spectral density that scales in the low frequency limit as $P(f) \sim f^{-\beta}$. 
The exponent $\beta$ is related to the DFA exponent by the scaling relation $\beta = 2\alpha - 1$ \cite{Heneghan2000,Kiyono2015}.
In exercise physiology, the power spectrum of heart rate time series is a frequently employed tool to quantify the cardiological response to exercise.
However, analyses in the frequency domain are potentially plagued by non-stationarity. For stationary signals, the autocorrelation function $C(\tau)=\langle X(\tau_0) X(\tau_0+\tau)\rangle$ decays for long lags $\tau$ with a power law $\sim \tau^{-\gamma}$. Then the  scaling relation  $\gamma = 2 - 2\alpha$ holds \cite{Holl2015}. For more details on the DFA method and its relation to correlation functions, see Supplementary Appendix~\ref{sec:app_a}.

\section{Partial Autocorrelation Function (PACF)\label{sec:autocorrelation}}

It is instructive to supplement DFA analyses by a direct study of correlations at different time scales by computing the  autocorrelation function $C(\tau)$  and the {\it partial} autocorrelation function $\cC(\tau)$ at lag $\tau$.  The latter has been successfully used to identify the best autoregressive (AR) process to fit a time series, using the fact that $\cC(\tau)=0$ for all $\tau>p$ for an AR model of order $p$ \cite{BoxJenkins}. 
The autocorrelation function $C(\tau)$ is dominated by trends in the data, suggesting apparent correlations. On the contrary, $\cC(\tau)$ is less affected by trends due to the subtraction of the linear dependence on intermediate lags from the autocorrelation function. 
If the time series contains oscillations, $C(\tau)$ shows a periodic pattern with a frequency that modulates the data. On the contrary, $\cC(\tau)$ shows anticorrelations. The lags $\tau$ for which $\cC(\tau)$ assumes negative values can be related to the periodicity of the oscillations.
More specifically, for a time series $X(\tau)$ the partial autocorrelation function is given by
\begin{equation}
\cC(\tau)=\left\langle [X(\tau_0)-\hat{X}_{\tau_0\tau}(\tau_0)][X(\tau_0+\tau)-\hat{X}_{\tau_0\tau}(\tau_0
+\tau)]\right\rangle
\label{pacf}
\end{equation}
where $\hat{X}_{\tau_0\tau}(\tau')$ is the best linear predictor, determined by
\begin{equation}
\hat{X}_{\tau_0\tau}(\tau') = c_0+ \sum_{i=1}^{\tau-1}c_i X(\tau_0+i) \, , 
\end{equation}
where the coefficients $c_i$ are determined by the conditions
\begin{align}
c_0 + \sum_{i=1}^{\tau-1} c_i \left\langle  X(\tau_0+i)  \right\rangle & = 
\langle X(\tau') \rangle \, ,\\
c_0 \left\langle  X(j) \right\rangle+ \sum_{i=1}^{\tau-1} c_i \left\langle  X(\tau_0+i)  X(j)\right\rangle & = 
\langle X(\tau')X(j) \rangle
\end{align}
for $j=\tau_0+1,\ldots, \tau_0+\tau-1$.
The function $\cC(\tau)$ can be computed practically from the Yule-Walker equations  \cite{BoxJenkins}.
The relation between the AR fits and $\cC(\tau)$ is useful in the performed analysis. Indeed, it has been shown that a signal with non-trivial periodic behavior can be, for short time scales, successfully fitted by an AR process, and its dominant frequency of oscillation can be extracted from the estimated coefficients. These findings have been also confirmed by DFA \cite{MeyerKantz2019}.

\section{Additional remarks on DFA for the validation of DDFA}
\label{sec:app_a}
In this section we provide some known theoretical results for the conventional DFA algorithm to support the validation procedure presented in the following section.
In DFA, the range of detectable exponents is determined by the degree of detrending, $n$, and is given by $0 \leq \alpha \leq n+1$ \cite{Kiyono2015}. While the existence of values $\alpha > 1$ may be criticized as a failure of the detrending procedure \cite{Bryce2012}, they may also be understood as an advantage of the method for allowing meaningful quantification of non-stationary processes \cite{Holl2016,Holl2019}. The detrending may be considered successful if it achieves the statistical equivalence over the DFA windows, so that the fluctuation function $F(s)$ does not depend on the window \cite{Holl2016,Holl2019}. This condition is fulfilled for DFA-$n$ with time series exhibiting polynomial trends of degree $n-1$. In general, for two {\it uncorrelated} signals  $X_A(t)$, $X_B(t)$ (random processes or trends), a superposition principle holds, stating that 
the squared fluctuation function of the sum $X_{A+B}(t) = X_A(t) + X_B(t)$ is given by $ F_{A+B}^2(s)=  F_A^2(s) +  F_B^2(s)$ \cite{Hu2001}.

For stationary processes and for non-stationary processes with stationary increments the fluctuation function $F_s$ does not depend on the window and may be analytically computed \cite{Holl2015,Lovsletten2017}. Its squared value is determined as the weighted sum of the autocovariance function $\hat C(\tau) = \mean{X(\tau_0) X(\tau_0 + \tau) } - \mean{X}^2$ in the former case, and that of the variogram $S(\tau) = \mean{\left[X(\tau_0 + \tau) - X(\tau_0)\right]^2}$ in the latter case, 
\begin{align}
F_s^2 &= \sum_{j=-s+1}^{s-1} G(j,s) \hat C(j)
\label{eq:autocovariance fluctuation function} \\
F_s^2 &= -\sum_{j=1}^{s-1} G(j,s) S(j)
\label{eq:variogram fluctuation function}
\end{align}
with the weight function $G(j,s)$ given by \footnote{We adapt the notation from Ref.~\cite{Lovsletten2017} with the following changes: The factor $s^{-1}$ is included in the weight function $G(j,s)$ that is symmetrically extended for negative $j$ by $G(-j,s) = G(j,s)$. This allows for a more succinct notation for Eqs.~\ref{eq:autocovariance fluctuation function} and \ref{eq:variogram fluctuation function}.}
\begin{align}
G(j,s) &= \frac{1}{s} \sum_{k=1}^{s-\abs{j}} a_{k,k+\abs{j}} \text{,}
\end{align}
where $a_{k,k'}$ are the elements of the matrix
\begin{align}
\mat{A} = \mat{D}\transpose \left[\mat{I} - \mat{B}\transpose \left(\mat{B}\mat{B}\transpose\right)^{-1}\mat{B}\right]\mat{D} \text{,}
\end{align}
where the elements $d_{i,j}$ of the matrix $\mat{D}$ are unity if $i \geq j$ and zero otherwise \cite{Holl2015,Lovsletten2017}. The effect of detrending is incorporated into the so-called design matrix $\mat{B}$ of least squares regression, which for DFA-1 is  given by \cite{Lovsletten2017}
\begin{align}
\mat{B} &= \begin{bmatrix}
1 & 1 & \cdots & 1 \\
1 & 2 & \cdots & s \\
\end{bmatrix} \, .
\end{align}
This matrix $\mat{A}$ describes an operator for constructing the squared fluctuations $F_{s}^2 = \mat{X}_{s,w}\transpose\mat{A}\mat{X}_{s,w}$ from the values of the time series $\mat{X}_{s,w}$ in window $w$ at the scale $s$ \cite{Lovsletten2017}.

The autocovariance function for fractional Gaussian noise (fGn) and the variogram for fractional Brownian motion (fBm) with Hurst parameter $H$ are known to be
\begin{align}
\hat C_H^{\text{fGn}}(\tau) &= \frac{\sigma^2}{2} \left( \abs{\tau + 1}^{2H} - 2\abs{\tau}^{2H} + \abs{\tau - 1}^{2H} \right)\, , \label{eq:fgn autocovariance} \\
S_H^{\text{fBm}}(\tau) &= \sigma^2 \abs{\tau}^{2H} \label{eq:fbm variogram} \text{,}
\end{align}
where $\sigma^2$ corresponds to the variance of ordinary Gaussian noise \cite{Mandelbrot1968}.
From these correlations and by using Eqs.~\ref{eq:autocovariance fluctuation function} and \ref{eq:variogram fluctuation function}, the theoretical fluctuation function may be computed for these processes. The spectrum of the scaling exponent $\alpha(s)$ is then obtained from \eqref{eq:local alpha}. These theoretical results may be utilized for studying the behavior of the DFA method as a function of the scale $s$. For example, the deviation between the DFA estimate for $\alpha(s)$ and the asymptotic large scale exponent $\alpha$ is visualized in Supplementary Fig.~\ref{fig:fgn fbm alpha deviation} for fGn and fBm. The well-known overestimation of the scaling exponent at the shortest scales is clearly visible, and it is most pronounced in the anticorrelated region with $\alpha < \nicefrac{1}{2}$. Around the asymptotic value of $\alpha=1$ there is an abrupt qualitative change as the scaling exponent is suddenly underestimated for an extended range of scales. This has been observed previously \cite{Lovsletten2017}. 
\begin{figure}
    \includegraphics{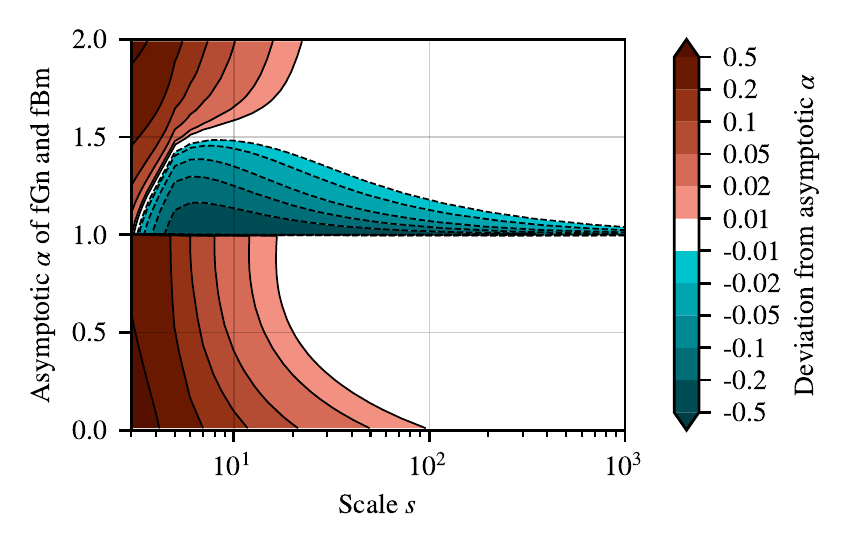}
    \caption{\label{fig:fgn fbm alpha deviation}Theoretical deviation of the DFA estimate $\alpha(s)$ from asymptotic scaling exponent $\alpha$ for fGn and fBm as a function of the scale $s$. The deviation is defined as the theoretical scale-dependent exponent $\alpha(s)$ minus the asymptotic scaling exponent $\alpha$. Note the quasi-logarithmic scale for the deviation.}
\end{figure}

In general, the short scale behavior depends on the details of the underlying process, and hence can be different for other processes such as  an autoregressive model AR(p).  For more details on the scale dependence of the deviation between asymptotic $\alpha$ and $\alpha(s)$, please see also Ref.~\cite{Molkkari:2018uf}.

\section{Numerical Validation of DDFA \label{sec:validation of the methods}}

Fractional Brownian motion (fBm) and its increments, fractional Gaussian noise (fGn), are commonly utilized for benchmarking DFA (see, e.g., Ref.~\cite{Shao2012} and references therein).
These processes are characterized by the Hurst parameter $0 \leq H < 1$, and exhibit long-range correlations with the asymptotic (large-scale) scaling exponents $\alpha = H$ for fGn and $\alpha = H + 1$ for fBm \cite{Mandelbrot1968,Lovsletten2017}.
Deviations from the asymptotic behavior occur at shorter scales due to the finite length of the samples and the intrinsic bias in DFA due to the detrending.
It is, however, possible to compute the exact theoretical scale-dependent scaling exponent $\alpha(s)$ for these processes, as described in Supplementary Appendix~\ref{sec:app_a}.

We validate the dynamic DFA (DDFA) method by applying it to simulated fGn and fBm, and comparing the results to the theoretically expected values. We utilize the Davies--Harte method, which is an efficient method for simulating these processes with their exact covariance structure \cite{Davies1987}\footnote{The accuracy of the simulated time series is established by comparing their joint fluctuation functions (the mean in \eqref{eq:dfa fluctuation function} is taken over the squared fluctuations $F_{s,w}^2$ of all realizations of the simulated time series) to the theoretical fluctuation functions of fGn and fBm.}.
We generate $10^3$ samples of fGn and fBm of length $10^5$ for each value of the Hurst parameter $H$. From these simulated time series, we compute the dynamic scaling exponent $\alpha(t,s)$ in non-overlapping dynamic segments with various dynamic segment length factors $\dlf$. 
The mean difference between the DDFA exponent $\alpha(t,s)$ and the theoretically expected DFA exponent $\alpha(s)$ is illustrated in Supplementary Fig.~\ref{fig:dynamic dfa bias mean}.
The general trend is that the limited sample size results in underestimation of the scaling exponents: the shorter the dynamic segments, the greater the bias. Similarly, processes with larger asymptotic $\alpha$ suffer from larger bias, with a weak discontinuity at $\alpha=1$ (when the process changes from fGn to fBm). This tendency can be understood as arising from the increased abundance and length of streaks in more correlated time series. This effect is  not fully captured by the relatively short segments. However, for the shortest scales (5 and 6), particularly in the anticorrelated region, the exponent is slightly overestimated instead. We also observe that the contour lines for the bias in $\alpha(t,s)$ have nearly converged to a constant value of the asymptotic $\alpha$ already at the scale $s = 40$.
\begin{figure}
    \includegraphics{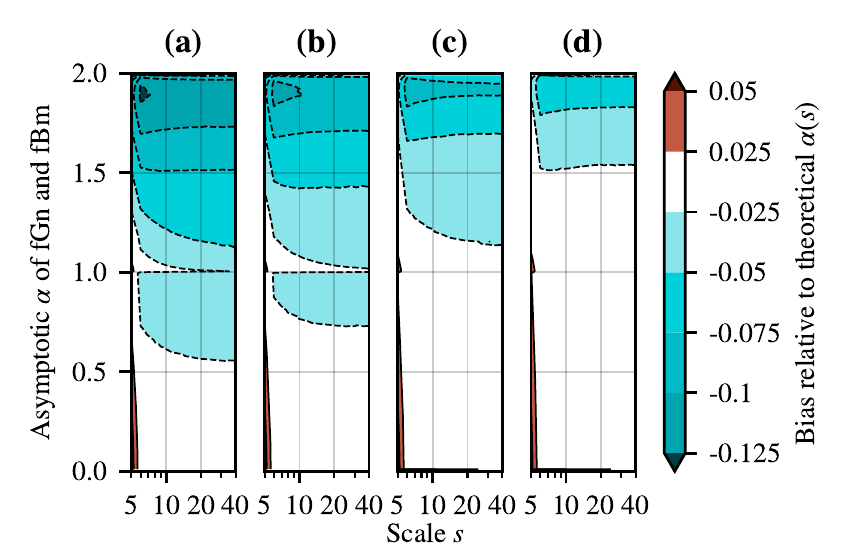}
    \caption{\label{fig:dynamic dfa bias mean} Bias of the dynamic detrended fluctuation analysis (DDFA) estimate of $\alpha(t,s)$ relative to its theoretically expected value for fractional Gaussian noise (fGn) and fractional Brownian motion (fBm). Characterized by the Hurst parameter $0 \leq H < 1$, the asymptotic $\alpha$ for these processes is $\alpha=H$ and $\alpha=H+1$ respectively. Details for computing the theoretical scale-dependent DFA exponent $\alpha(s)$ are provided in Supplementary Appendix~\ref{sec:app_a}. The bias is defined as the observed $\alpha(t,s)$ minus its theoretically expected value $\alpha(s)$. The dynamic segment length factors $\dlf$ are 4, 5, 7, and 10 in (a-d), respectively.}
\end{figure}

The standard deviation of the DDFA estimation of the exponent $\alpha$ is shown in Supplementary Fig.~\ref{fig:dynamic dfa bias std} for segment lengths 4, 5, 7, and 10 in (a-d), respectively. The deviation consistently decreases with the increasing segment length. On the other hand, the deviation reduced as the DFA-1 exponents 
approach the limits $\alpha=0$ and $\alpha=2$, since at these boundaries there is less room for variations. In particular, as $\alpha$ is reduced from $1/2$ towards
zero (corresponding to the anticorrelated regime) the deviations are strongly reduced. The local reduction in the deviation just above $\alpha=1$ is due to
the most anticorrelated increments in this regime (see Supplementary Table~\ref{tb:dfa scaling exponent}).
\begin{figure}
    \includegraphics{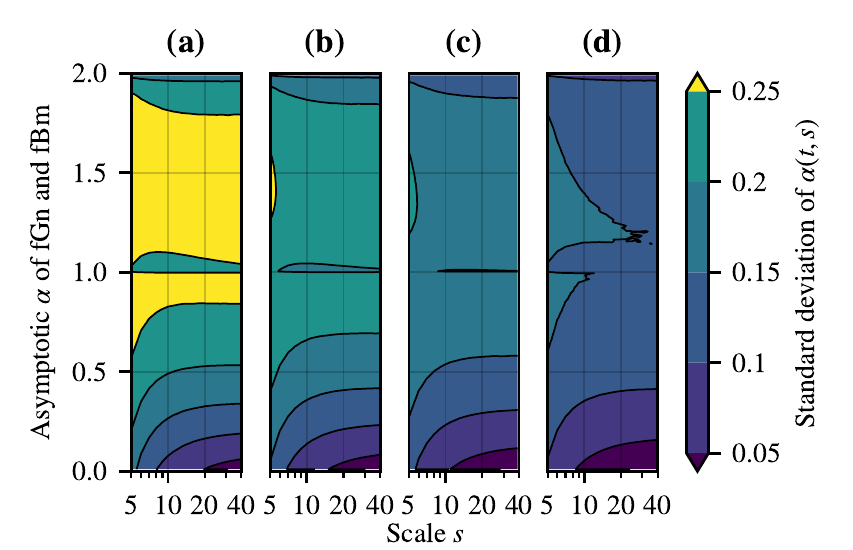}
    \caption{\label{fig:dynamic dfa bias std} Standard deviation of the dynamic detrended fluctuation analysis (DDFA) estimate of $\alpha(t,s)$ for fGn and fBm as a function of the scale $s$. Characterized by the Hurst parameter $0 \leq H < 1$, the asymptotic $\alpha$ for these processes is $\alpha=H$ and $\alpha=H+1$ respectively. The dynamic segment length factors $\dlf$ are 4, 5, 7, and 10 in (a-d), respectively.}
\end{figure}

Generally, the bias and the standard deviation of the DDFA method are found to be acceptable for our purposes, especially in view of the fact that we have particular
interest in the anticorrelated region as underlined below in the results. All our DDFA computations are performed with the dynamic segment length factor $\dlf = 5$. This was found to be a good compromise between the accuracy of the DDFA method and the dynamical resolution requiring a sufficiently small segment size.

\section{Additional Heartbeat Correlation Plots}
\label{app:plots}
Here we present beat-to-beat (RR) interval (RRI) correlations for all the subjects in the study. In Supplementary Fig.~\ref{fig:mit bpm aggregate} we illustrate the average RRI correlation results as a function of the heart rate aggregated over all the runs for each subject of Group T. The relative heart rate is utilized to better facilitate the comparison between different individuals. Similar correlation plots for the marathons of Group M are shown in Supplementary Fig.~\ref{fig:marathons overview}, along with the correlation landscapes as a function of time during the marathon races. Additionally, we establish the consistency of the anticorrelated bands in the presence of possible trends in Supplementary Fig.~\ref{fig:results trend consistency}. We demonstrate this by limiting the analysis to subsets of data where the heart rate within the dynamic segments exhibits subsequently lower and lower standard deviation. The analysis is performed for subject T07, who has the most data.

Each correlation plot consists of pairs of color-coded DDFA (upper panels) and DPACF (lower panels) results.
Plots as a function of the heart rate (HR) are based on data that is averaged of dynamic segments whose average HR falls within bins with widths of \num{0.1} BPM or \num{0.001} for the absolute and relative HR, respectively. The values for empty bins are linearly interpolated if the gap does not exceed \num{0.5} BPM (absolute) or \num{0.005} (relative). The DDFA plots (as a function of the HR) also display the conventional short-scale (4--16 RRIs) scaling exponents $\alpha_1$ by a semi-transparent black line with error bars (thin bars: standard deviation, thick bars: standard error of the mean, barely visible). The exponent is computed in moving windows of 50 RRIs in HR bins of \num{2} BPM (absolute) or \num{0.01} (relative). The DDFA plots displaying the correlation landscapes for single runs show the instantaneous HR with the semi-transparent black line instead, and in the corresponding single run DPACF plots the values that do not pass the non-zero significance test are shown in white.

\begin{figure*}
    \includegraphics{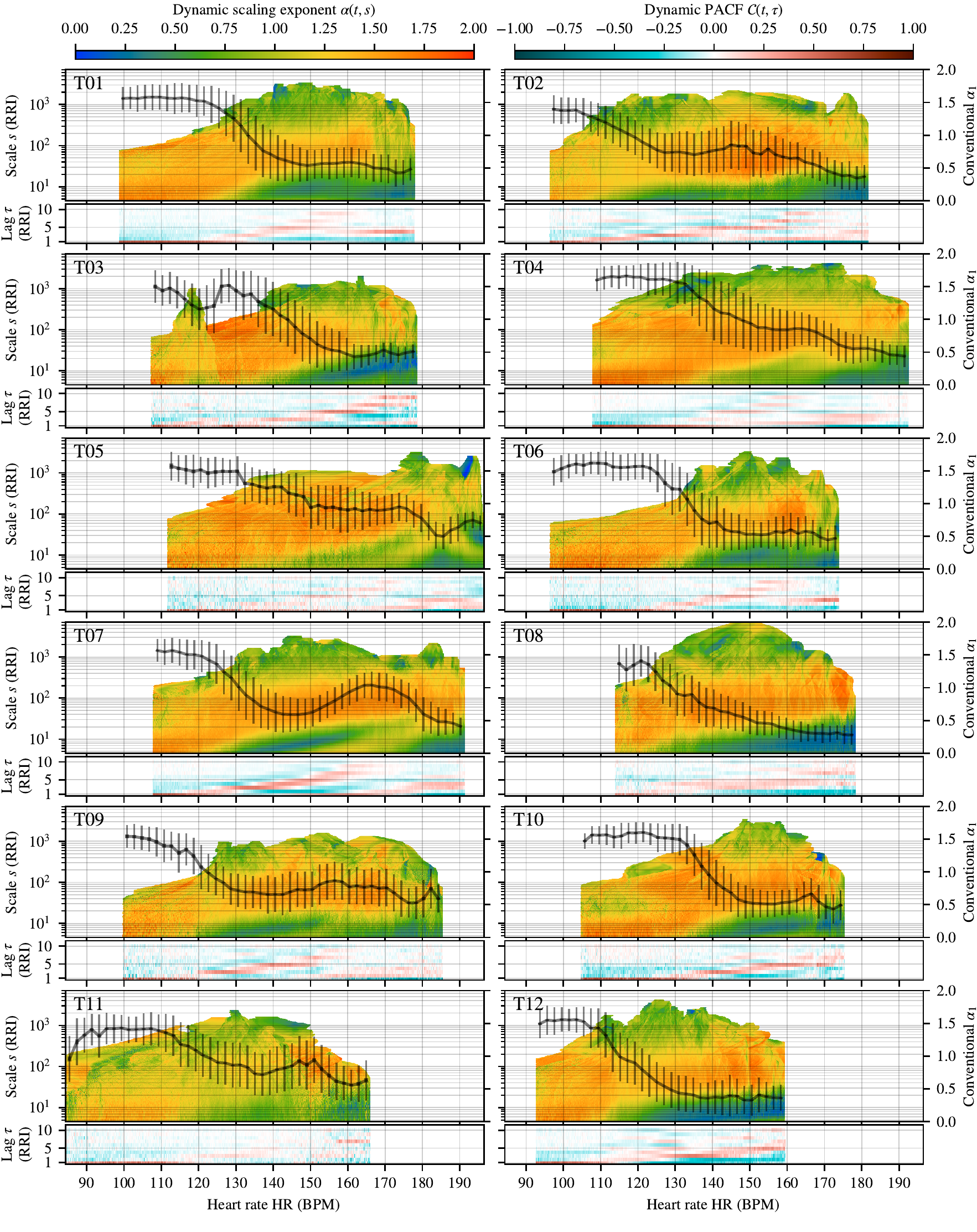}
    \caption{\label{fig:mit bpm aggregate}Aggregate RRI correlation results for each subject of Group T. For each subject the average DDFA-1 scaling exponents $\alpha(t,s)$ (upper panels) and DPACF-0 correlations $\mathcal{C}(t,\tau)$ (lower panels) as a function of binned relative heart rate.
    For a detailed explanation about how the data is computed, please see Supplementary Appendix~\ref{app:plots}.}
\end{figure*}
\begin{figure*}
    \includegraphics{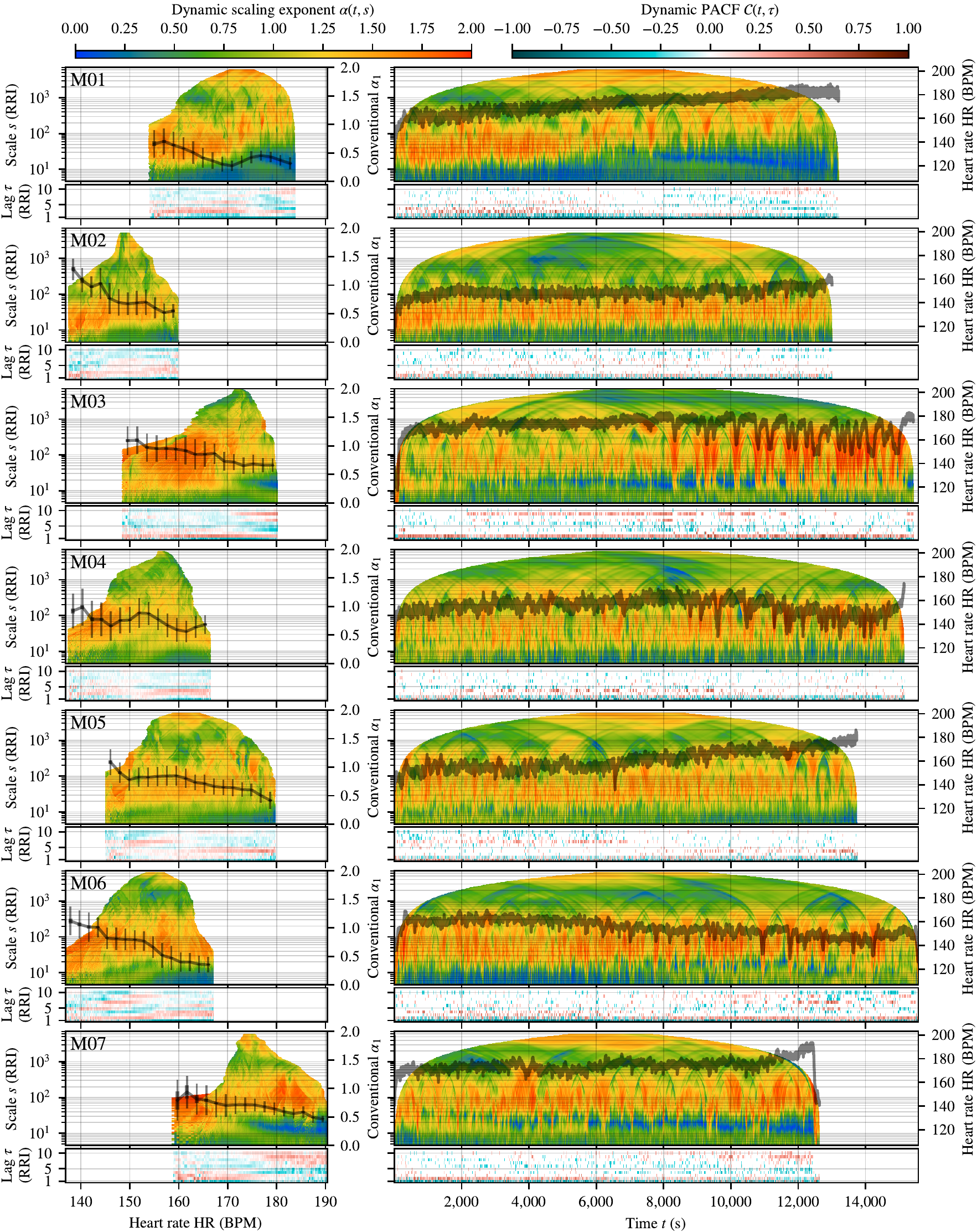}
    \caption{\label{fig:marathons overview}Overview of all the marathons in Group M. Left: Average RRI correlation results as a function of binned relative heart rate for each subject of Group M. Right: RRI correlation landscapes of the marathon runs. For each subject the DDFA-1 scaling exponents $\alpha(t,s)$ (upper panels) and DPACF-0 correlations $\mathcal{C}(t,\tau)$ (lower panels) are shown.
    For a detailed explanation about how the data is computed, please see Supplementary Appendix~\ref{app:plots}.}
\end{figure*}
\begin{figure*}
    \includegraphics{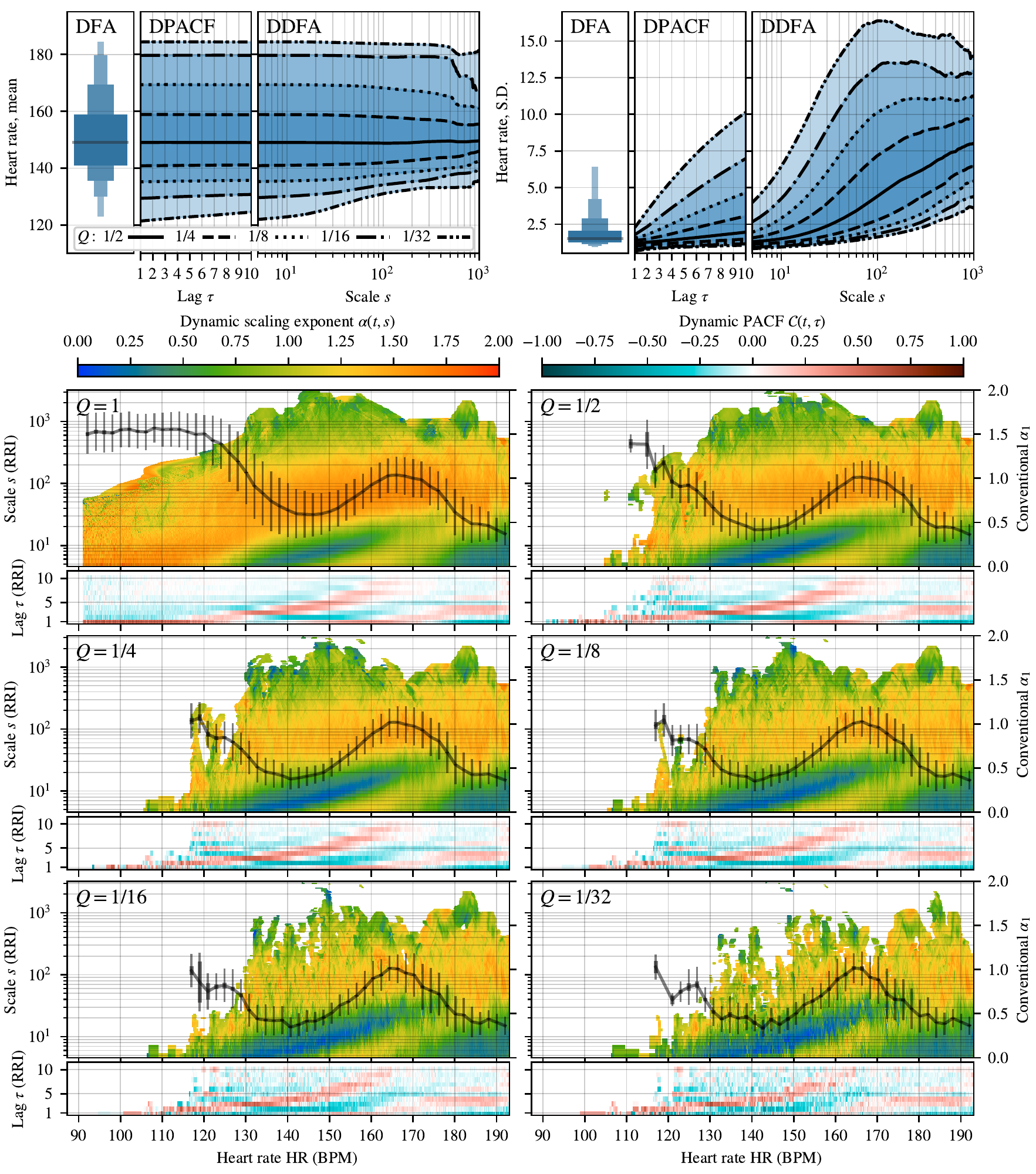}
    \caption{\label{fig:results trend consistency}Consistency of the results with respect to trends. Top row: different quantiles $Q$ of the mean and standard deviation of the heart rate within the dynamic segments for all the data of subject T07. For DPACF-0 and DDFA-1 the dynamic segment length factor $a$ has values of 10 and 5, respectively, and for conventional DFA the short-range (4--16 beats) scaling exponent is computed in moving windows consisting of 50 RRIs. Lower panels: the average RRI correlation results as a function of the heart rate when the data is limited to dynamic segments with the heart rate standard deviation less than the value for the specified quantiles $Q$. For a detailed explanation about how the data is computed, please see Supplementary Appendix~\ref{app:plots}.}
\end{figure*} 
\FloatBarrier

\bibliography{bibliography}

\end{document}